\documentclass[10pt]{iopart}

\expandafter\let\csname equation*\endcsname\relax
\expandafter\let\csname endequation*\endcsname\relax

\usepackage{amsmath, amsbsy}
\usepackage[T1]{fontenc}
\usepackage{xurl}
\usepackage{cite}
\usepackage[unicode=true, bookmarks=true, bookmarksnumbered=true, bookmarksopen=true, bookmarksopenlevel=1, breaklinks=false,pdfborder={0 0 0}, backref=false, colorlinks=true]{hyperref}
\hypersetup{citecolor=blue,linkcolor=blue,urlcolor=blue}

\usepackage[dvipsnames]{xcolor}
\usepackage{lipsum,multicol}

\usepackage[english]{babel}
\usepackage[autostyle, english = american]{csquotes}
\MakeOuterQuote{"}

\usepackage{graphicx}
\usepackage{subcaption}
\newcommand{\labelphantom}[1]{%
{\phantomsubcaption%
\label{#1}}%
}%
\makeatletter
\renewcommand\p@subfigure{\thefigure-}
\makeatother

\newcommand{\sub}[2]{#1_\mathrm{#2}}

\begin{document}

\title[Options for heating, fueling, and current drive in MANTA, a high-field NT fusion pilot plant]{Options for heating, fueling, and current drive in MANTA, a high-field negative triangularity fusion pilot plant}
\author{%
H.~S.~Wilson$^{1,2,\ast}$,
A.~O.~Nelson$^{1}$,
J.~M.~Park$^{3}$,
E.~Hassan$^{3,4}$,
C.~Collins$^{3}$,
and
C.~Paz-Soldan$^{1}$
}

\address{$^{1}$~Columbia University, New York, NY 10027, USA}
\address{$^{2}$~MIT Plasma Science and Fusion Center, Cambridge, MA 02139, USA}
\address{$^{3}$~Oak Ridge National Lab, Oak Ridge, TN 37830, USA}
\address{$^{4}$ Physics, Faculty of Science, Ain Shams University, Cairo 11566, EG}
\address{}
\address{\textit{$^{\ast}$ Corresponding author:} hsw2119@columbia.edu}

\vspace{10pt}
\begin{indented}
\item[]\today
\end{indented}

\begin{abstract}
In this work, we assess how the performance of MANTA, a high-field, highly radiative, negative triangularity tokamak concept, depends on pellet fueling, ion cyclotron resonant frequency (ICRF) heating, and lower hybrid current drive (LHCD) schemes. Negative triangularity (NT) has been gaining momentum in the fusion community as a naturally, robustly ELM-free regime that is capable of power plant level performance. The MANTA device has been the subject of previous works regarding the whole-device design, power handling and divertor assessments, and core transport optimizations. Informed by these previous studies, we assess whether changes to the original MANTA heating and current drive scheme can positively affect performance, both in terms of lowering scrape-off layer power (improving power-handling, a primary benefit of NT) and maximizing plasma gain to levels necessary for an attractive fusion pilot plant. In this work, we find that the MANTA concept can yield improved performance in terms of plasma gain and scrape-off layer power by simultaneously going to lower input power and higher seeded impurity fraction. Also, we find that increasing the fueling magnitude can increase fusion power while minimally impacting scrape-off layer power. Performance can also be improved by optimizing the ICRF design; decreasing the minority ion fraction, decreasing the toroidal mode number of the injected wave, and switching from a helium-3 minority ion to a hydrogen-1 minority ion all result in increases to fusion power. The additional inboard space afforded by the NT geometry provides an opportunity for wave launch to be moved to the inboard side. This is beneficial for LHCD access and results in a wide range of potential current profiles in the edge, with the total driven current due to 25 MW of LHCD reaching up to 1.2 MA, or about 12\% of the total current.
\end{abstract}

\submitto{\PPCF}
\maketitle
\ioptwocol

\section{Introduction}
\label{sec:intro}

The dominant strategy for a tokamak fusion power plant (FPP) is to operate in positive triangularity (PT) high-confinement mode (H-mode) \cite{butteryAdvancedTokamakPath2021,fableSelectedTransportStudies2016,hillesheimOverviewPhysicsBasis2026,kimDesignUpdatesAnalysis2020}. This choice is motivated by the increased confinement in PT H-mode and by the wealth of experiments and models that have been used to validate its core physics. PT H-mode exhibits an instability called an edge-localized mode (ELM) that results in large energy fluence to plasma-facing components \cite{keilhackerConfinementStudiesHtype1984,leonardEdgelocalizedmodesTokamaks2014}. While the largest (type-I ELMs) are tolerable in present machines (though they still can cause impurities to enter the plasma from interacting with the plasma-facing components \cite{duxMainChamberSources2011,vanrooijTungstenDivertorErosion2013}), existing experiments operate in regimes that differ significantly from those that will be required for a net-power device. At the higher current, pressure, magnetic field, and power of a net-power fusion device, type-I ELMs will be detrimental to the device and must be avoided at all costs \cite{gunnSurfaceHeatLoads2017,eichELMDivertorPeak2017}.

Experiments on DIII-D \cite{Austin2019,marinoniDivertedNegativeTriangularity2021,nelsonRobustAvoidanceEdgeLocalized2023,paz-soldanSimultaneousAccessHigh2024}, AUG \cite{happelOverviewInitialNegative2022,auconeExperimentsModellingNegative2024}, TCV \cite{codaEnhancedConfinementDiverted2022,pochelonEnergyConfinementMHD1999}, and MAST-U \cite{nelsonFirstAccessELMfree2024} as well as high fidelity core modeling \cite{merloNonlocalEffectsNegative2021,marianiFirstprincipleBasedPredictions2024,digiannataleSystemSizeScaling2024,bielajewGyrokineticProfilePrediction2025} have all shown that negative triangularity (NT) plasmas can robustly exhibit H-mode-like confinement without ELMs. Additionally, NT plasmas do not require the scrape-off layer power ($\sub{P}{SOL}$) to remain above a certain threshold to maintain high confinement as in PT H-mode. This allows for the use of techniques to reduce $\sub{P}{SOL}$ (such as seeding high-Z noble gas impurities \cite{luxImpurityRadiationDEMO2015,bernertPowerExhaustSOL2017,casaliFirstIntegrationNegative2023}) and thus ease the challenge of divertor integration. From a power-handling standpoint, the lack of ELMs in NT plasmas with high confinement, the absence of a requirement to maintain the scrape-off layer power above the L-H transition power, and the naturally larger divertor wetted area enabled by the NT shape all make NT an attractive FPP regime. Recent NT experiments have shown that NT has appealing performance in normalized parameters that scale to FPPs \cite{paz-soldanSimultaneousAccessHigh2024} and provides a clear, robust path to improved power-handling over PT H-mode scenarios. 

Given the high cost of building a fusion device, performance prediction using computational models is a necessary precursor to the fusion power plant. A computational test-bed enables relatively low-cost forecasting of the sensitivity of performance to key parameters (that serve as proxies for tunable experimental parameters) before they are enacted in an actual fusion environment. A few computational studies of varying fidelity exist that explore the advantages of NT as an FPP regime \cite{kikuchiNegativeTriangularityTokamak2014,Medvedev2015}, but only two extend to the whole-device; one uses lower fidelity codes with surrogate models to scan a broad number of potential builds \cite{slendebroekExploringFusionPower2026}, while the other is a higher-fidelity analysis of a single operating point, named MANTA \cite{collaborationMANTANegativetriangularityNASEMcompliant2024,millerPowerHandlingHighlyradiative2024}, where further studies have documented the performance dependence on key core parameter assumptions\cite{wilsonCharacterizingNegativeTriangularity2024}. However, none of the aforementioned NT FPP studies used a heating and current drive (H\&CD) model that was both high-fidelity and self-consistently integrated into the plasma core simulation workflow. In this work, we focus on the MANTA design point, as the parameters for its successful operation are clearly outlined in \cite{collaborationMANTANegativetriangularityNASEMcompliant2024}. MANTA's NT geometry affords previously inaccessible inboard space that could be beneficially leveraged for enhanced H\&CD in the design. To our knowledge, this is the first dedicated study of the effect of H\&CD on the performance of an NT FPP.

Sources of H\&CD vary, but a few dominate in most present experiments. For example, most tokamaks have current primarily driven by the central solenoid and are primarily heated with neutral beam injection. However, certain technologies that are relied upon for the successful operation of present devices may not work as effectively in the conditions of a net-power fusion device. Most current FPP designs (both NT and PT) plan to use some form of resonant frequency (RF) H\&CD. Electron cyclotron (EC) RF H\&CD is efficient, can be launched from outside the vacuum vessel, and doesn't require the inclusion of a minority heating ion that would dilute the plasma. However, it may not work effectively in high-field devices because even the most cutting-edge gyrotrons produce waves with FPP-relevant power at a frequency of no more than 236 GHz  \cite{shinyaDemonstrationHighpowerGyrotron2026}. This frequency of launched wave would correspond to a resonance layer at a magnetic field of about 8.4 T from the equation 

\begin{equation}
\sub{B}{res}=2\pi f\sub{m}{e}/e
\end{equation}

where $\sub{B}{res}$ is the magnetic field at which resonance occurs for a given launched frequency $f$, $\sub{m}{e}$ is the mass of an electron, and $e$ is the elementary charge. On a device like MANTA, the resonance layer would be located outside of the plasma because the field decays as 1/$R$. Ion cyclotron (IC) RF waves, on the other hand, can readily damp on-axis even in a high-field device. However, ICRF does not drive current very efficiently and requires a minority ion species in the plasma \cite{stixFASTWAVEHEATINGTWOCOMPONENT,frankSimulatingEnergeticIons2024}. Lower hybrid (LH) RF current drive is less straightforward, as the lower hybrid frequency depends on both the magnetic field and the density. Therefore, the angle and parallel index of refraction of the launched wave must be chosen carefully for the wave to damp fully inside the plasma. LHCD is very efficient at driving current off-axis and thus can be used effectively for stabilizing the plasma. It is of note that both ICRF and LHCD require the use of in-vessel antennas. Due to the high heat loads projected in an FPP, this adds engineering complexity that an external driver would not have. The bootstrap current in a plasma does not rely upon direct external drivers but is an intrinsic current arising from the pressure gradient in a tokamak \cite{kesselBOOTSTRAPCURRENTTOKAMAK}, and thus maximizing the bootstrap current fraction in an FPP is desirable \cite{butteryAdvancedTokamakPath2021}. A study on DIII-D showed that for similar operational parameters in a PT and NT shot (plasma current, toroidal magnetic field, and auxiliary power), bootstrap current was similar for both cases. However, in NT it is distributed as a broader profile across the plasma radius, whereas PT had a large amount in the edge due to the steep pressure of the pedestal \cite{marinoniHmodeGradeConfinement2019}.

\begin{figure}
    \centering
    \includegraphics[width=\linewidth]{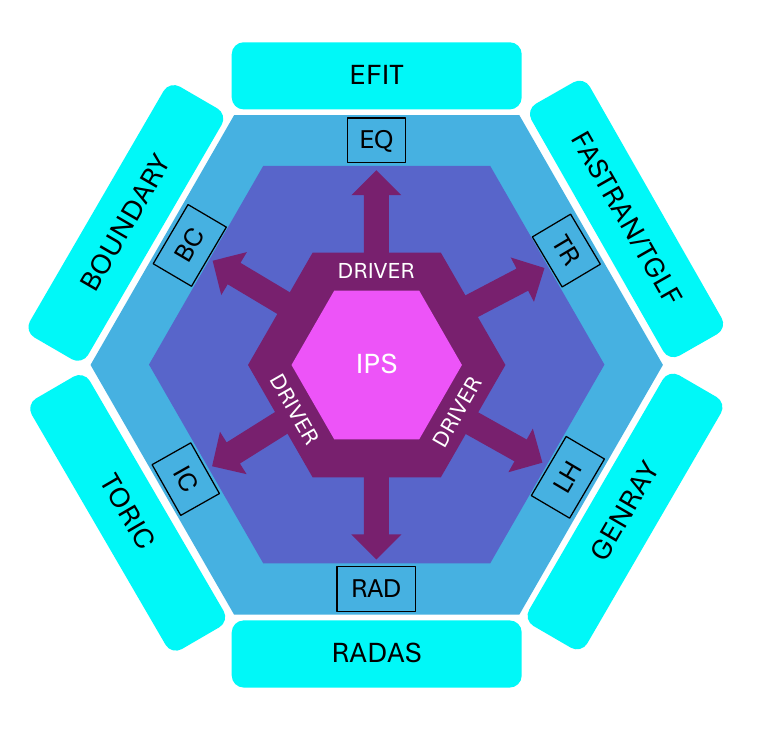}
    \caption{A diagram representing the capabilities of the IPS-FASTRAN code utilized in this work. EQ is for equilibrium, TR is for transport, LH is for lower hybrid, RAD is for radiation, IC is for ion cyclotron, and BC is for boundary condition. Figure modified from \cite{hassanCorePedestalPlasmaConfigurations2023}.}
    \label{fig:IPS_diagram}
\end{figure}

This paper is organized as follows. In section \ref{sec:methods}, we discuss the motivation behind using the NT fusion pilot plant design MANTA as a starting point and describe the integrated modeling framework used throughout this work. In section \ref{sec:basecase}, a base case is established incorporating evolution across a larger simulation region, self-consistently evolving H\&CD, and including intrinsic impurities. Considerations for a particle source for density evolution and its effect on performance are covered in section \ref{sec:density_evolution}. In section \ref{sec:pauxscan}, we show that the performance of our base case can be optimized by lowering the input power magnitude, using full simulations that include line radiation, temperature evolution, density evolution, and self-consistent heating calculations. We then discuss options for the optimization of an ICRF source in section \ref{sec:ic}, including the choice of minority ion and the fraction of minority ion and their effect on antenna performance and effective power deposition. In section \ref{sec:lh}, we establish an effective range for the potential current profiles for the high-field side launch of LH resonant waves in the base case. Conclusions are presented in section~\ref{sec:conclusion}.


\section{Simulation setup and methods}
\label{sec:methods}

\begin{figure*}
    \centering
    \labelphantom{fig:basecase-a}
    \labelphantom{fig:basecase-b}
    \labelphantom{fig:basecase-c}
    \labelphantom{fig:basecase-d}
    \includegraphics[width=0.49\linewidth]{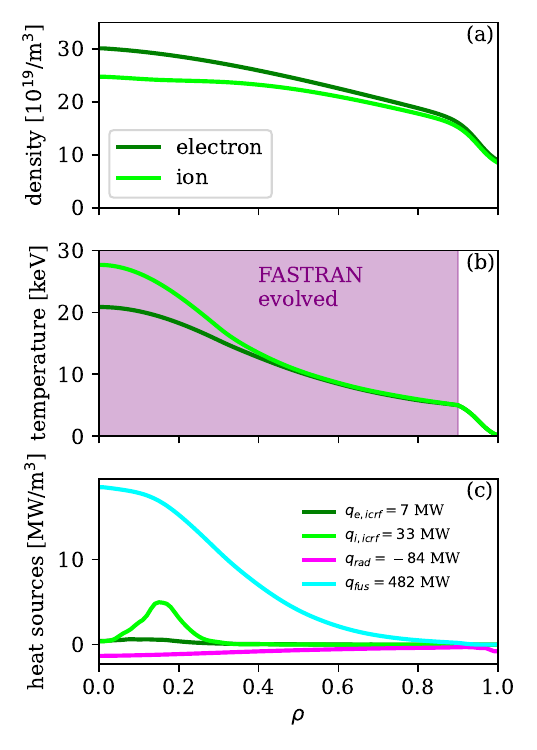}
    \includegraphics[width=0.49\linewidth]{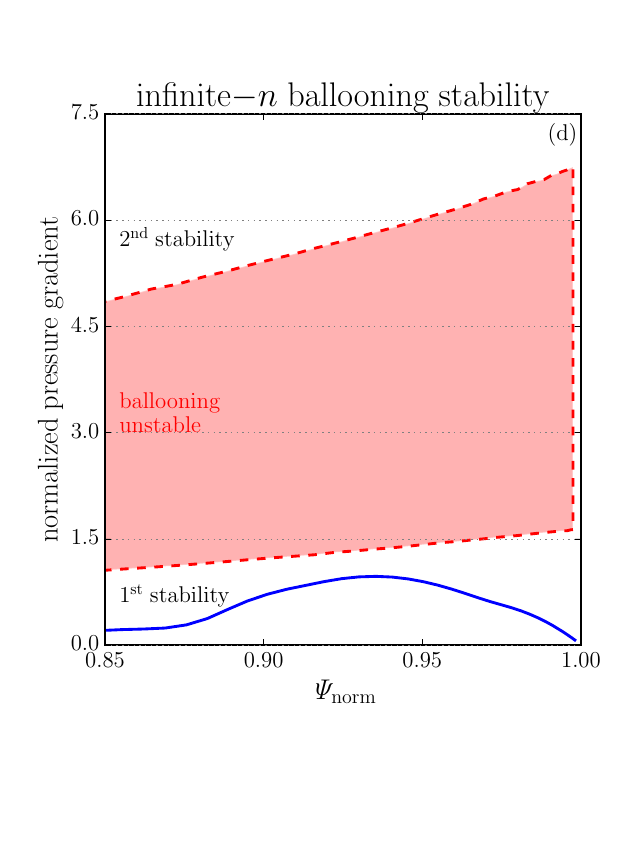}
    \caption{Selected profiles versus normalized radius $\rho$ are shown for the base case used for the studies in this work. Electron and ion quantities are in dark green and light green, respectively. Plot (a) shows density, plot (b) shows temperature, and plot (c) shows selected heat sources and sinks, all against $\rho$. The temperature plot has shaded the region where the ion and electron temperatures are evolved in FASTRAN. Heat source plot shows the power density profiles for ICRF heating of electrons ($\sub{q}{e,icrf}$), ICRF heating of ions ($\sub{q}{i,icrf})$, total radiative loss ($\sub{q}{rad})$, and total fusion heating $\sub{q}{fus}$, with corresponding colors indicated in the legend. Plot (d) is an infinite-$n$ ballooning stability diagram, with the normalized pressure gradient of the base case in blue, and plotted against normalized poloidal flux $\sub{\Psi}{norm}$. The ballooning unstable region is shaded red.}
    \label{fig:basecase}
\end{figure*}

\subsection{Base MANTA equilibrium and profiles}
\label{subsec:basecase}

While the insights gained from this work are intended to be broadly applicable to NT FPP design, we focus primarily on the MANTA operating point due to a few factors. First, a dedicated study of H\&CD in an NT FPP does not exist in the present literature. MANTA is a good candidate for this study as it is a whole-device design for an NT FPP that extends from the plasma core to the cryostat surrounding its high-temperature superconducting (HTS) magnets, and meets or exceeds all suggested parameters of performance for an FPP outlined in the National Academies of Science, Engineering, and Medicine's report "Bringing Fusion to the U.S. Grid" \cite{NAP25991}. Its use of HTS magnets to produce an 11 T field on-axis presents unique considerations for H\&CD that up to this point have only been explored for a handful of designs \cite{linPhysicsBasisICRF2020, sorbomARCCompactHighfield2015, hillesheimOverviewPhysicsBasis2026}. Also, there are two follow-up studies that have been done on its core physics \cite{wilsonCharacterizingNegativeTriangularity2024} and power-handling \cite{millerPowerHandlingHighlyradiative2024} facets, which provide further information on what aspects of the design are integral or optional for its success. In reference \cite{wilsonCharacterizingNegativeTriangularity2024}, a "MANTA-like" base case was established with higher density and temperature boundary conditions that remains infinite-$n$ ballooning stable and reaches an H-factor of near 1 ($\sub{H}{98y2}$ from the IPB98(y,2) scaling in \cite{confinementChapter2Plasma1999}) and a Greenwald fraction of near 1 ($\sub{f}{Gr}$ from the scaling in \cite{greenwaldDensityLimitsToroidal2002}). NT experiments have shown that these values of normalized confinement and normalized density are readily accessible \cite{paz-soldanSimultaneousAccessHigh2024a,sauterOperationGreenwaldDensity2025}. Thus, all simulations in this work start with the equilibrium and profiles from the $\sub{H}{98y2}\approx 1$ and $\sub{f}{Gr} \approx 1$ base case (designated in this paper as MANTA-1) unless otherwise noted. The details of MANTA-1 are outlined in table \ref{tab:MANTA_vs_new}.

\subsection{Integrated modeling with IPS-FASTRAN}
\label{subsec:ipsfastran}

All codes used in this work are implemented as part of the Integrated Plasma Scenario (IPS) framework \cite{usdoeofficeofsciencescIntegratedPlasmaSimulator2011,elwasifDesignImplementationSwim2010,foleyIntegratedFusionPlasmaSimulation2008,HPCSimToolsIPSframework2026}. IPS allows for the self-consistent integration of various external models of plasma equilibrium, core transport, heating, current drive, and stability to find a steady state solution. The version we are using includes the 1-D radial core transport solver FASTRAN \cite{parkEfficientTransportSolver2017} that predicts density and temperature profiles using fluxes calculated by TGLF \cite{staeblerTheorybasedTransportModel2007}, a quasilinear turbulent transport code. FASTRAN uses a high-order numerical scheme that solves for each variable as well as its gradient and is able to predict accurate profiles more rapidly by reducing the number of calls to TGLF. TGLF has been shown in previous studies to reproduce NT experiments well on DIII-D \cite{mcclenaghanExaminingTransportIntegrated2021}, TCV \cite{balestriExperimentsGyrokineticSimulations2024}, and AUG \cite{auconeExperimentsModellingNegative2024}. In references \cite{balestriExperimentsGyrokineticSimulations2024} and \cite{auconeExperimentsModellingNegative2024}, TGLF was implemented with the SAT2 saturation rule. However, in reference \cite{mcclenaghanExaminingTransportIntegrated2021}, it was found that SAT0 and SAT2 predicted the DIII-D NT profiles with similar accuracy. We will implement TGLF with SAT0 in this work to more easily compare with the results in reference \cite{wilsonCharacterizingNegativeTriangularity2024} (where SAT0 was used in simulations that evolved particle and heat flux simultaneously) and with the results in reference \cite{slendebroekExploringFusionPower2026} (where SAT0 was used for the transport calculations in a comparative analysis of PT and NT FPP-class tokamaks). 

IPS-FASTRAN has been validated on existing tokamak experiments, such as DIII-D \cite{parkIntegratedModelingHigh2018,butteryDIIIDResearchPrepare2019,holcombSteadyStateScenario2014} and KSTAR \cite{APS60thAnnual,yoonKSTARProgramClosing2015}. It has also been used for performance prediction of conceptual machines such as ITER \cite{pettyAdvancesSteadystateHybrid2017,murakamiIntegratedModellingSteadystate2011}, K-DEMO \cite{kangDevelopmentSystematicSelfconsistent2017}, and CAT \cite{butteryAdvancedTokamakPath2021}. A simple diagram of the IPS-FASTRAN capabilities used in this work is shown in figure \ref{fig:IPS_diagram}. The IPS-FASTRAN framework employs a lightweight component architecture written in Python, where the outputs of existing codes are shared between modules using a centralized data file \cite{parkIntegratedModelingHigh2018}. The modules we use include EFIT \cite{laoReconstructionCurrentProfile1985} for equilibrium construction, TGLF/FASTRAN for turbulent transport calculations, GENRAY \cite{meneghiniIntegratedModelingApplications2015} for lower hybrid resonance H\&CD, the RADAS model \cite{bodyRADAS2025} for line radiation of impurity species, and the full-wave code TORIC \cite{brambillaNumericalSimulationIon1999} for ion cyclotron resonance H\&CD. The "BOUNDARY" module can be implemented within the framework to assert boundary conditions for the initial current profile of each simulation. The Chang-Hinton model \cite{changEffectFiniteAspect1982} is implemented for neoclassical transport calculations, while the Sauter model \cite{sauterNeoclassicalConductivityBootstrap1999} is implemented for bootstrap current calculations. 

\section{Implementing MANTA in IPS-FASTRAN}
\label{sec:basecase}
Throughout this work, IPS-FASTRAN provides the capability to self-consistently integrate higher fidelity H\&CD modules into evaluations of the plasma state. The temperature and density profiles of a plasma can influence the efficiency and deposition location of resonant H\&CD methods, so it is important to adjust predictions based on the results of profile evolution in FASTRAN to find a more accurate steady-state solution. There are a few other advantages gained from using IPS-FASTRAN over the tools used in the original MANTA work and subsequent core studies in \cite{wilsonCharacterizingNegativeTriangularity2024} that increase the fidelity of our model. First, the use of FASTRAN to expedite profile predictions by evaluating TGLF at a smaller number of grid points means evolution across a larger region (up to normalized radius $\rho =  0.9$ in this case) is less computationally expensive. In \cite{collaborationMANTANegativetriangularityNASEMcompliant2024}, TGYRO/TGLF was evolved to $\rho = 0.85$ and in \cite{wilsonCharacterizingNegativeTriangularity2024}, only to $\rho = 0.8$. This is significant because reference \cite{wilsonCharacterizingNegativeTriangularity2024} found that the predicted kinetic profiles for an FPP-class NT tokamak depend heavily on the boundary condition chosen for the edge. Evolving to $\rho = 0.9$ for NT simulations follows work done in \cite{slendebroekExploringFusionPower2026} of predicting NT FPP performance for a range of designs. 

It is of note that the pressure gradient in the region from $\rho = 0.9$ to $1.0$ has been shown in NT experiments on DIII-D to be limited by the infinite-$n$ ballooning stability \cite{nelsonCharacterizationELMfreeNegative2024}. In the original MANTA design as well as in the base case in this work, the normalized pressure gradient is well below the infinite-$n$ ballooning instability region. In our base case, it is at $\sim 72$\% the limit at the peak pressure gradient. Thus, there may be room to increase the pressure boundary condition beyond what is presented here, which could result in a significant increase in performance \cite{wilsonCharacterizingNegativeTriangularity2024}. It is of note that in predictions of the edge pressure in PT H-mode, the commonly-used EPED model \cite{snyderFirstprinciplesPredictiveModel2011a} assumes the pressure gradient grows to meet the full peeling-ballooning stability limit. Reference \cite{slendebroekExploringFusionPower2026} estimates the profile shape in the region from $\rho = 0.9$ to $1$ partially by use of an experimentally observed relation between stored energy in the core ($\rho = 0$ to $0.9$) and edge ($\rho = 0.9$ to $1$) regions in NT plasmas on DIII-D.  In our base case, stored energy in the edge is slightly higher than the range of the value predicted by this relation ($\sub{W}{th,ped}/\sub{W}{th,core} = 0.38$ here while $0.30 \pm 0.05$ is predicted), but it is of note that we did not implement the same edge interpolation scheme as in reference \cite{slendebroekExploringFusionPower2026}. For this work, we will use the boundary conditions of the "MANTA-like" base case (here referenced as MANTA-1) in reference \cite{wilsonCharacterizingNegativeTriangularity2024}, but we note changes to performance due to the edge boundary condition of an NT FPP are well explored in \cite{wilsonCharacterizingNegativeTriangularity2024}.

\begin{table}
\begin{center}
    \begin{tabular}{c|c|c}
             & MANTA-1 & IPS base case \\
        \hline\hline $\sub{B}{t}$ (T) & 11    & 11      \\
        \hline $\sub{I}{P}$ (MA)& 10    & 10       \\
        \hline $\sub{R}{maj}$ (m) & 4.6 & 4.6     \\
        \hline $\sub{a}{minor}$ (m) & 1.2 & 1.2    \\
        \hline $\delta$         & -0.5  & -0.5     \\
        \hline $\kappa$         & 1.4   & 1.4      \\
        \hline $\sub{\beta}{N}$ &       1.9      &          1.5    \\
        \hline $\sub{q}{95}$    &     2.6        &       2.0       \\
        \hline $\sub{P}{aux}$ (MW)  & 40   & 40       \\
        \hline $\sub{P}{fus}$ (MW)  &  966  &    482     \\
        \hline $\sub{Q}{plasma}$  &  24  &   12  \\
        \hline $\sub{P}{SOL}$ (MW) & 91 &     52       \\
        \hline $\sub{P}{rad}$ (MW) &  142 &     84        \\
        \hline $\sub{f}{Kr}$ & 0.001 & 0.0008 \\
        \hline $\sub{H}{98y2}$ &  1.0 &  0.9                 \\
        \hline $\sub{f}{Gr}$ & 0.94 &  0.91                   \\
        \hline $\langle{n}\rangle$ ($10^{19}/m^3$) & 41  &   39   \\
        \hline $\langle\sub{T}{e}\rangle$ (keV) & 9.4  &    8.1    \\
        \hline $\langle\sub{T}{i}\rangle$ (keV) & 9.1  &    8.6    \\
        \hline $\sub{n}{e,0.8}$ ($10^{19}/m^3$) & 18 &  19\\
        \hline $\sub{T}{e,0.8}$ (keV) & 6.8 & 5.8       \\
    
    \end{tabular}
    \caption{Parameters for the MANTA-like core scenario \cite{wilsonCharacterizingNegativeTriangularity2024} and changes by implementation into the IPS-FASTRAN framework. Brackets indicate volume average.}
    \label{tab:MANTA_vs_new}
    \end{center}
\end{table}

An additional benefit of FASTRAN is that it facilitates more rapid density profile construction from TGLF calculations of particle flux, so our simulations in this work include density evolution, whereas previous work relied on scaling laws to estimate density peaking. We also include more impurities in the simulations in this work. Depending on the study, this may include tungsten to account for wall sources, a minority ion for use with ion cyclotron resonant heating (discussed in section \ref{sec:ic}), helium-4 from fusion reactions in the core, and krypton for radiating power in the core to facilitate divertor integration. The inclusion of these impurities results in fuel dilution and thus lowers the resultant fusion power output.

Kinetic electron and ion temperature and density profiles for the base case in this work are shown in figure \ref{fig:basecase}. The electron temperature and density profiles were initialized to match those in \cite{wilsonCharacterizingNegativeTriangularity2024}.  A ballooning stability diagram for the edge is also shown in plot \ref{fig:basecase-d}, where the normalized pressure gradient is defined as 
\begin{equation}
    \alpha=\frac{\sub{\mu}{0}}{2\pi^2}\frac{\partial V}{\partial \psi}\left(\frac{V}{2\pi^2\sub{R}{maj}}\right)^{1/2}\frac{\mathrm{d}p}{\mathrm{d}\psi}
\end{equation} 
where $V$ is the volume enclosed by each flux surface, $\psi$  is the poloidal flux, and $p$ is the total plasma pressure. Thus, plot \ref{fig:basecase-d} demonstrates that the edge interpolation of the profiles results in the normalized pressure gradient remaining in the first stability region. This follows the expectation of infinite-$n$ ballooning stability in an NT edge, where access to the second stability region is closed and thus the plasma does not enter H-mode \cite{nelsonCharacterizationELMfreeNegative2024}.  The electron density is held constant and the profile peaking matches that predicted by the scaling in \cite{angioniScalingDensityPeaking2007}. The ion density was calculated assuming quasineutrality with tungsten at a constant fraction of the electron density of $1.5\times10^{-5}$, helium-3 at 0.02, krypton at 0.0008, and helium-4 calculated from fusion reactions in the core, which accounts for the dip in ion density seen in the core. The electron and ion temperatures were then evolved independently in FASTRAN using TGLF with a heating profile calculated by TORIC. Dilution in the core results in a lower fusion power and lower radiated power than seen in reference \cite{wilsonCharacterizingNegativeTriangularity2024}. We were unable to converge a case with a higher krypton fraction to increase the radiated power without altering the boundary condition of the profiles or heating scheme due to radiative collapse. In the case shown in figure \ref{fig:basecase}, the plasma gain $\sub{Q}{plasma}$ is 12 and the scrape-off layer power $\sub{P}{SOL}$ is 52 MW. Additional parameters for this base case are shown in table \ref{tab:MANTA_vs_new} for comparison to the MANTA-1 core scenario.

There are multiple factors that lead to the differences in performance between the two cases. First, the boundary condition in the IPS base case is set to be equal to that of MANTA-1 at $\rho = 0.9$, leading to the lower temperature value at $\rho = 0.8$ ($\sub{T}{e,0.8} = 5.8$ keV) and the higher density at $\rho = 0.8$ ($\sub{n}{e,0.8} = 19 \times 10^{19}/m^3$) because transport in the IPS base case includes the region from $\rho = 0.8$ to $0.9$. The change in temperature boundary condition is enough to significantly lower the peaking of the evolved temperature profile, which follows results shown in reference \cite{wilsonCharacterizingNegativeTriangularity2024}. The electron density profile is held constant, but including tungsten, helium-3, and helium-4 in the simulation led to a decrease in the ion density due to dilution from maintaining quasineutrality. Both the decreased average temperature and density contribute to decreased $\sub{P}{fus},$ and $\sub{\beta}{N}$. A lower $\sub{P}{fus}$ in the IPS base case does result in a lower $\sub{P}{SOL}$ according to 
\begin{equation}
\sub{P}{SOL} = \sub{P}{alpha} + \sub{P}{aux} + \sub{P}{ohm} - \sub{P}{rad,core}
\end{equation}
from reference \cite{nelsonInterpretativeSOLModeling2021}, where $\sub{P}{ohm}$ is the ohmic power and $\sub{P}{alpha}$ is the alpha power, or $\sub{P}{fus}/5$ for a DT plasma. However, it is of note that $\sub{P}{rad,core}$ must be lowered further (to less than 40 MW) for the simple divertor design of MANTA \cite{collaborationMANTANegativetriangularityNASEMcompliant2024,millerPowerHandlingHighlyradiative2024}. A possibility to accomplish this with lower input power will be discussed in section \ref{sec:pauxscan}. Higher krypton fractions than 0.0008 in the IPS base case did not lead to convergence in the integrated IPS-FASTRAN workflow due to radiative collapse, while MANTA-1 was able to use a krypton fraction of 0.001. We suspect this is due to the boundary condition being moved to larger $\rho$ as well, because it is closer to the peak of the line radiation in the edge from krypton. 

In the remainder of this paper, we will leverage different H\&CD schemes and fueling models to increase the performance of this scenario, targeting $\sub{P}{fus} > 300$ MW and $\sub{P}{SOL} < 40$ MW, which are the parameters the MANTA design would need to meet the requirements for a fusion pilot plant set out in reference \cite{nationalacademyofengineeringBringingFusionUS2021} with a simple divertor solution. We note that in reference \cite{wilsonCharacterizingNegativeTriangularity2024}, it was found that changing the boundary condition at the edge of profile evolution (in this case at $\rho = 0.9$) can change performance significantly, and would thus be another route to improving the performance of this base case. However, this work intends to show how different heating schemes and fueling models change fusion performance, and so we will maintain the density and temperature boundary conditions at those shown in figure \ref{fig:basecase} unless otherwise noted. That is, this work does not exploit potential modifications to the boundary condition, which would result in additional performance changes, in order to isolate the effects on the plasma core from different H\&CD schemes. In reality, the boundary condition is set by physics in the edge region, which is out of the scope of this work. 

\begin{figure*}
    \centering
    \labelphantom{fig:pellet-a}
    \labelphantom{fig:pellet-b}
    \labelphantom{fig:pellet-c}
    \labelphantom{fig:pellet-d}
    \includegraphics[width=0.45\linewidth]{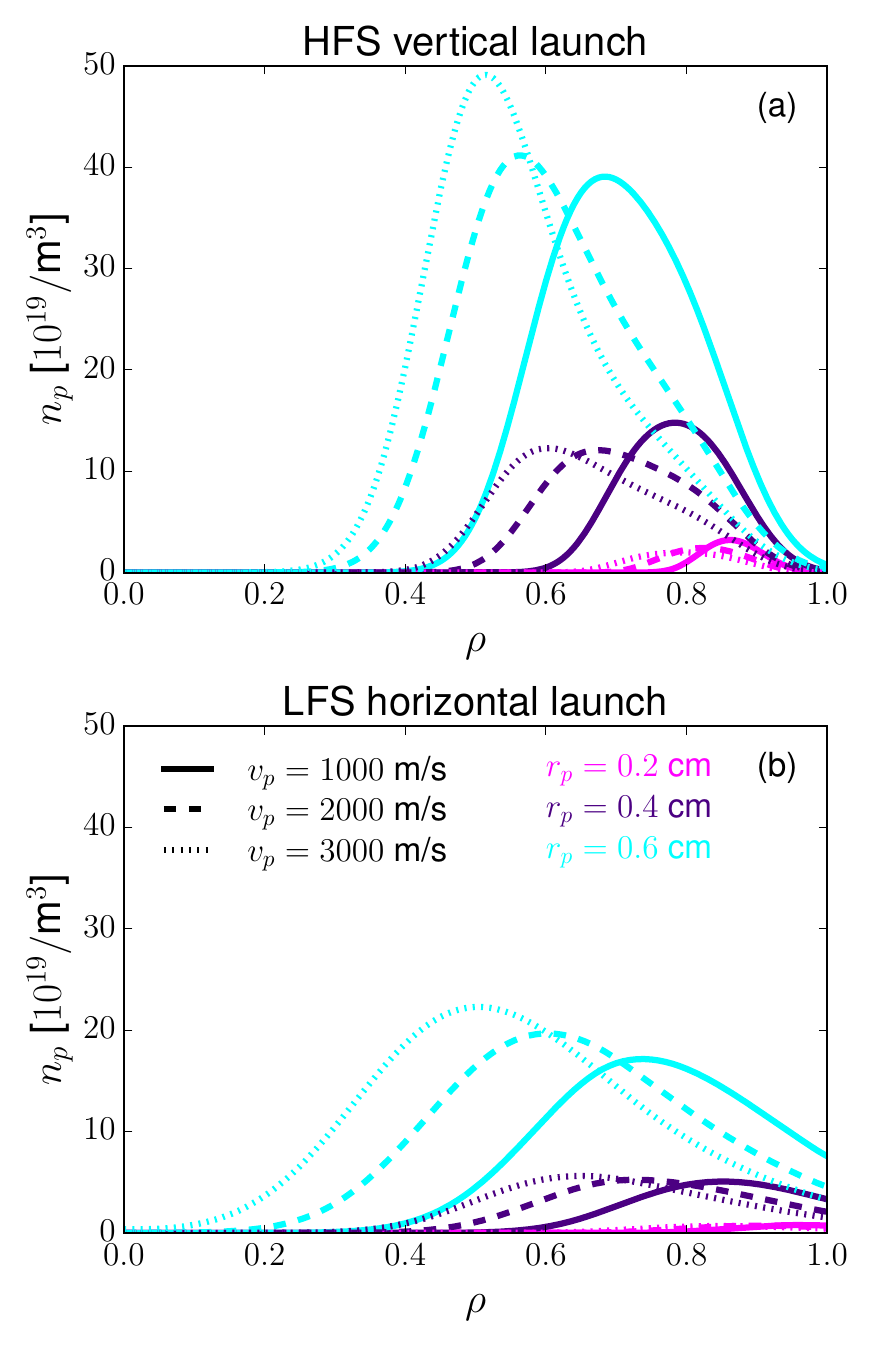}
    \includegraphics[width=0.45\linewidth]{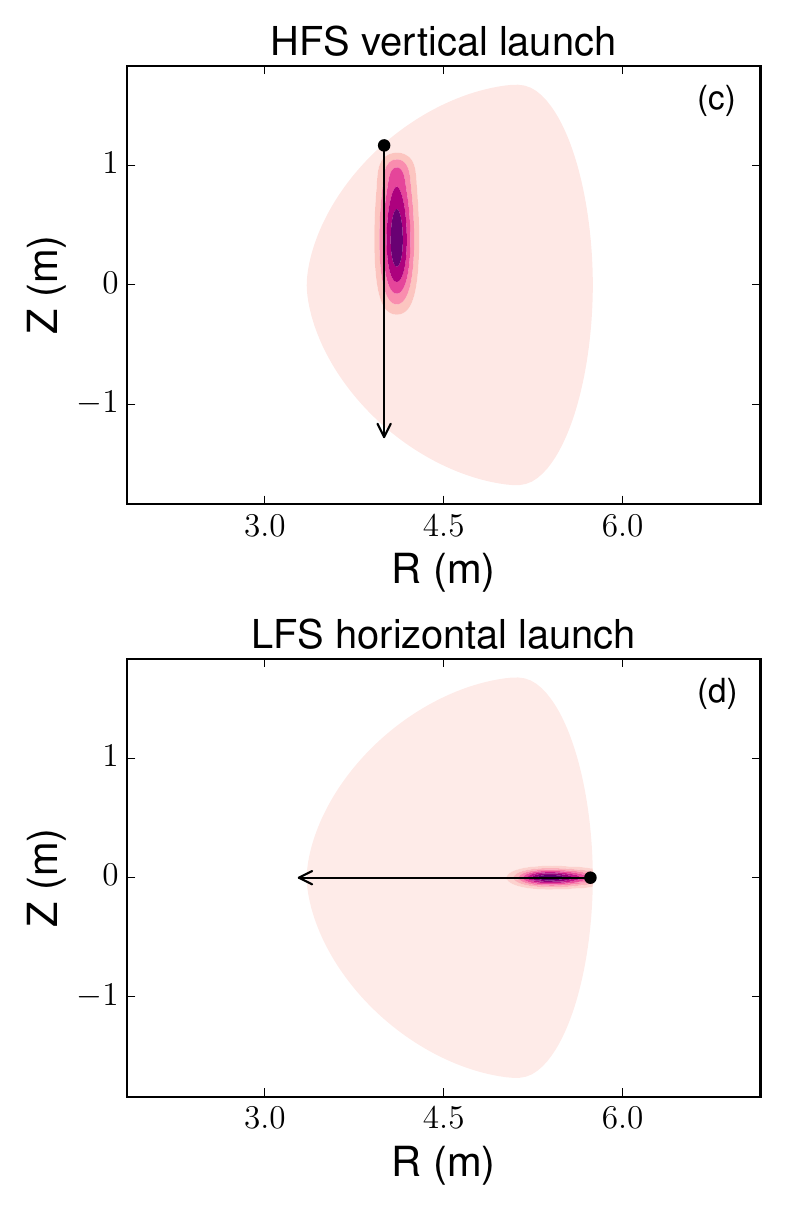}
    \caption{Pellet deposition for the MANTA scenario is shown. On the left plots, the 50/50 DT pellets range in radius ($\sub{r}{p}$) across 0.2 cm (pink), 0.4 cm (purple), and 0.6 cm (blue) and in injection speed ($\sub{v}{p}$) across 1000 m/s (solid line), 2000 m/s (dashed line), and 3000 m/s (dotted line). The right plots show 2D deposition for the $\sub{r}{p} = 0.6$ cm and $\sub{v}{p} = 3000$ m/s cases. High field side vertical launch is from R,Z = (4m, 1.17m), and low field side horizontal launch is from R,Z = (5.73m, 0m), both at the last closed flux surface and indicated by the black point. Arrows give the direction of the injection velocity.}
    \label{fig:pellet_deposition}
\end{figure*}

\section{Density evolution and fueling source assumptions}
\label{sec:density_evolution}

A challenge of evolving density in an FPP-class plasma simulation is the uncertainty of the fueling source. In a nuclear device like MANTA, neutral beam injection (NBI) may not be practical due to the size of necessary ports reducing the tritium breeding ratio (TBR) \cite{hemsworthConsiderationsDevelopmentNeutral2017,hopfNeutralBeamInjection2021}, which must be kept above unity. In MANTA, the TBR is 1.15, but $\sim10$\% of the volume of the breeding blanket would need to be removed for one ITER-class NBI port (1 m by 1.6 m \cite{hemsworthOverviewDesignITER2017}). Turning to other methods, one of the most promising for an FPP-class tokamak is pellet injection \cite{zhangSimulationStudyFueling2025,geulinPelletCoreFueling2022}. Deep penetration of pellets into a plasma is made more difficult by transit through a hot pedestal \cite{mcclenaghanSelfconsistentInvestigationDensity2023}, so the lack of a steep pedestal in NT should be beneficial for pellet deposition closer to the core. A common approximation of a pellet injection source is a Gaussian profile of an electron source located in the edge region, which is what we will employ in this work. Density evolution from a Gaussian source with flux matching done in FASTRAN yields increased fidelity over the assumption in previous MANTA work that density would follow the Angioni scaling \cite{angioniScalingDensityPeaking2007} for density peaking. In this section, we will cover the sensitivity of the performance of our base case to the choice of fueling source by scanning assumptions of the source Gaussian width $\sub{\Delta}{s}$, radial location $\sub{\rho}{s}$, and total integrated source particles per second $\sub{N}{s,tot}$.

To check the feasibility of the proposed particle deposition profiles from pellets, we use the simple and flexible Pellet Ablation Model (PAM) \cite{mcclenaghanSelfconsistentInvestigationDensity2023} found within the OMFIT suite of codes \cite{meneghiniIntegratedModelingApplications2015} to approximate possible particle deposition profiles from pellet injection in MANTA. Following work in reference \cite{mcclenaghanSelfconsistentInvestigationDensity2023} studying pellet injection in the FPP-class tokamak designs CFETR \cite{wanOverviewPresentProgress2017} and ITER, we implement a 50/50 DT pellet ranging in radius from 0.2 cm to 0.6 cm and injection speeds of up to 3000 m/s. A pellet speed of up to 3000 m/s is only possible with a direct-line-of-sight injection \cite{frattolilloInjectionHighspeedSolid2021}, as opposed to injection through the curved guide tubes often required to access the high field side \cite{geulinPelletCoreFueling2022}. High-field side pellet injection can take advantage of the beneficial $\nabla B$ drift effect discussed in reference \cite{mcclenaghanSelfconsistentInvestigationDensity2023} that moves the pellet mass in the major radius direction. The additional inboard space in NT enables easier integration of a straight guide tube to the high field side. The results of both high-field side and low-field side injection for MANTA are shown in figure \ref{fig:pellet_deposition}. For the HFS launch case, the pellet radii 0.2 cm, 0.4 cm, and 0.6 cm correspond to total integrated particles of about $0.2\times 10^{22}$, $1.7\times 10^{22}$, and $5.6\times 10^{22}$, respectively. For the LFS case, the total integrated particles range from $0.08\times 10^{22}$ to $4.2\times 10^{22}$. In plot \ref{fig:pellet-c}, the beneficial $\nabla B$ drift effect is shown, spreading the pellet mass toward the plasma center. When the injection location was moved to a larger major radius in the HFS case, it did little to improve pellet deposition profiles, as the pellet still did not travel all the way to the axis. Designing a pellet with a thin carbon shell could result in deeper penetration into the core because the carbon would take longer to ablate than the DT payload, but would also increase the $\sub{Z}{eff}$ of the plasma and result in some fuel dilution \cite{mcclenaghanSelfconsistentInvestigationDensity2023}. However, for our purposes, we are not attempting to optimize a fueling source but simply to increase confidence in a reasonably conservative fueling source for the remainder of the studies in this work. It is of note that the pellet ablation rate in this model scales as $\sub{B}{t}^{-0.35}$, so the pellet is traveling further before fully ablating due to the relatively high magnetic field in MANTA. However, this $\sub{B}{t}$ dependence has yet to be verified by experiment \cite{mcclenaghanSelfconsistentInvestigationDensity2023}.

\begin{figure}
    \centering
    \labelphantom{fig:peaking-a}
    \labelphantom{fig:peaking-b}
    \labelphantom{fig:peaking-c}
    \includegraphics[width=\linewidth]{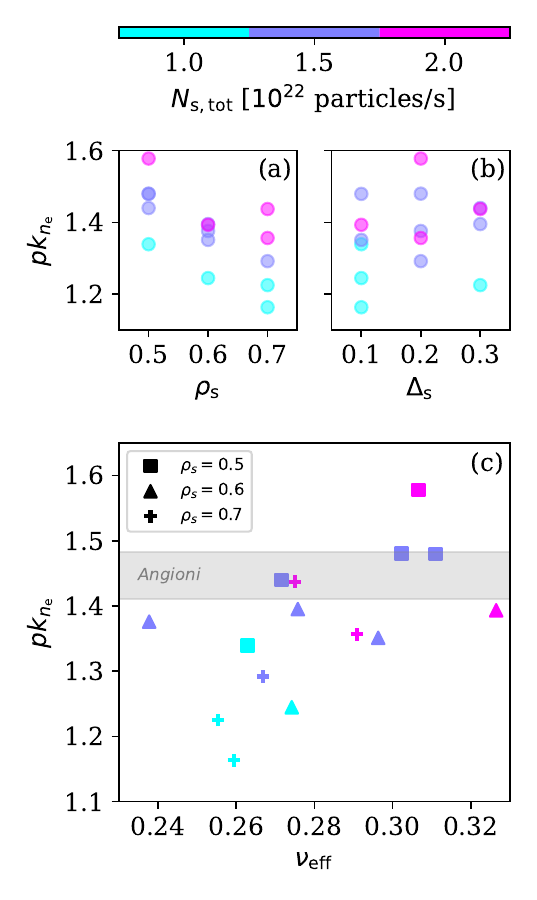}
    \caption{In these plots, each point is a converged simulation in a scan of fueling source shape and location. The color indicates $\sub{N}{s,tot}$ for the fueling source in each simulation. In (a) and (b), the density peaking $pk_{\sub{n}{e}}$ is plotted against $\sub{\rho}{s}$ and $\sub{\Delta}{s}$, respectively. In (c), $pk_{\sub{n}{e}}$ is plotted against $\sub{\nu}{eff}$, with data point shapes indicating $\sub{\rho}{s}$ of the fuel source used, given in the legend. The gray region shows the range of predicted density peaking values from equation \ref{eq:Angioni} for the simulations shown.}
    \label{fig:peaking}
\end{figure}

In figure \ref{fig:peaking}, three plots are shown from simulations scanning the shape of a Gaussian fueling source for evolving density in the base case, with profile shapes based on those in figure \ref{fig:pellet_deposition}. Each point represents a converged simulation, with electron density, electron temperature, and ion temperature evolved in IPS-FASTRAN, 40 MW of ICRF heating calculated self-consistently with TORIC, and line radiation from a 0.0008 fraction of krypton included. The color of each point corresponds to the total integrated particles per second $\sub{N}{s,tot}$ in the fuel source of the simulation. In plot \ref{fig:peaking-a}, density peaking $pk_{\sub{n}{e}} = \sub{n}{e}(\sub{\rho}{pol} = 0.2)/\sub{\langle \sub{n}{e} \rangle}{Vol}$ as defined in reference \cite{angioniScalingDensityPeaking2007} is plotted against $\sub{\rho}{s}$. For each color, it is clear that the $pk_{\sub{n}{e}}$ is increasing as the fuel source is moved inward (to smaller $\sub{\rho}{s}$). At each $\sub{\rho}{s}$, there is a clear trend of $pk_{\sub{n}{e}}$ increasing as $\sub{N}{s,tot}$ increases. Comparing $pk_{\sub{n}{e}}$ against $\sub{\Delta}{s}$ in plot \ref{fig:peaking-b} does not show a clear relation, leading us to draw the conclusion that the radial location of the source is more leveraging on the fusion power output than the width of the source in these ranges. This is likely due to the sources at larger values of $\sub{\rho}{s}$ needing to overcome more outward diffusion before achieving core peaking. 

In reference \cite{angioniScalingDensityPeaking2007}, it was found that variation in density peaking across the database studied could not be caused exclusively by the fueling (in their case, from neutral beam injection). Interestingly, the peaking of the density profiles in these simulations is very dependent on the shape and location of the fueling source, with the peaking increasing by up to 40\% of the minimum peaking. However, reference \cite{angioniScalingDensityPeaking2007}, the most statistically significant parameter in predicting density peaking was found to be collisionality. In the simulations shown in figure \ref{fig:peaking}, temperature is being evolved as well, which in turn affects collisionality and thus the density peaking. In plot \ref{fig:peaking-c}, $pk_{\sub{n}{e}}$ is plotted against effective collisionality ($\sub{\nu}{eff} = 0.2\langle \sub{n}{e} \rangle \sub{R}{maj}/\langle \sub{T}{e}\rangle^2$). Each plotted point is given a shape corresponding to the $\sub{\rho}{s}$ of the fueling source in the simulation, indicated in the legend. From this plot, there appears to be a positive correlation between $pk_{\sub{n}{e}}$ and $\sub{\nu}{eff}$, which is opposite to what is expected from scaling. Specifically, one Angioni predicted scaling is given by the equation
\begin{multline}
\label{eq:Angioni}
    \sub{pk}{Angioni} = 1.347 \pm 0.014 - (0.117 \pm 0.005)\log{(\sub{\nu}{eff})}\\
    + (1.331 \pm 0.117)\sub{\Gamma}{NBI}^* - (4.030 \pm 0.810)\beta
\end{multline}
copied from equation 3 in reference \cite{angioniScalingDensityPeaking2007} and thus predicts a negative correlation between $pk_{\sub{n}{e}}$ and $\sub{\nu}{eff}$. We note that in terms of the natural logarithm of $\sub{\nu}{eff}$, this data is over a very small range compared to the database in \cite{angioniScalingDensityPeaking2007}, so the apparent variation is not as significant. 

\begin{figure*}
    \centering
    \labelphantom{fig:pauxscan-a}
    \labelphantom{fig:pauxscan-b}
    \labelphantom{fig:pauxscan-c}
    \labelphantom{fig:pauxscan-d}
    \includegraphics[width=0.8\linewidth]{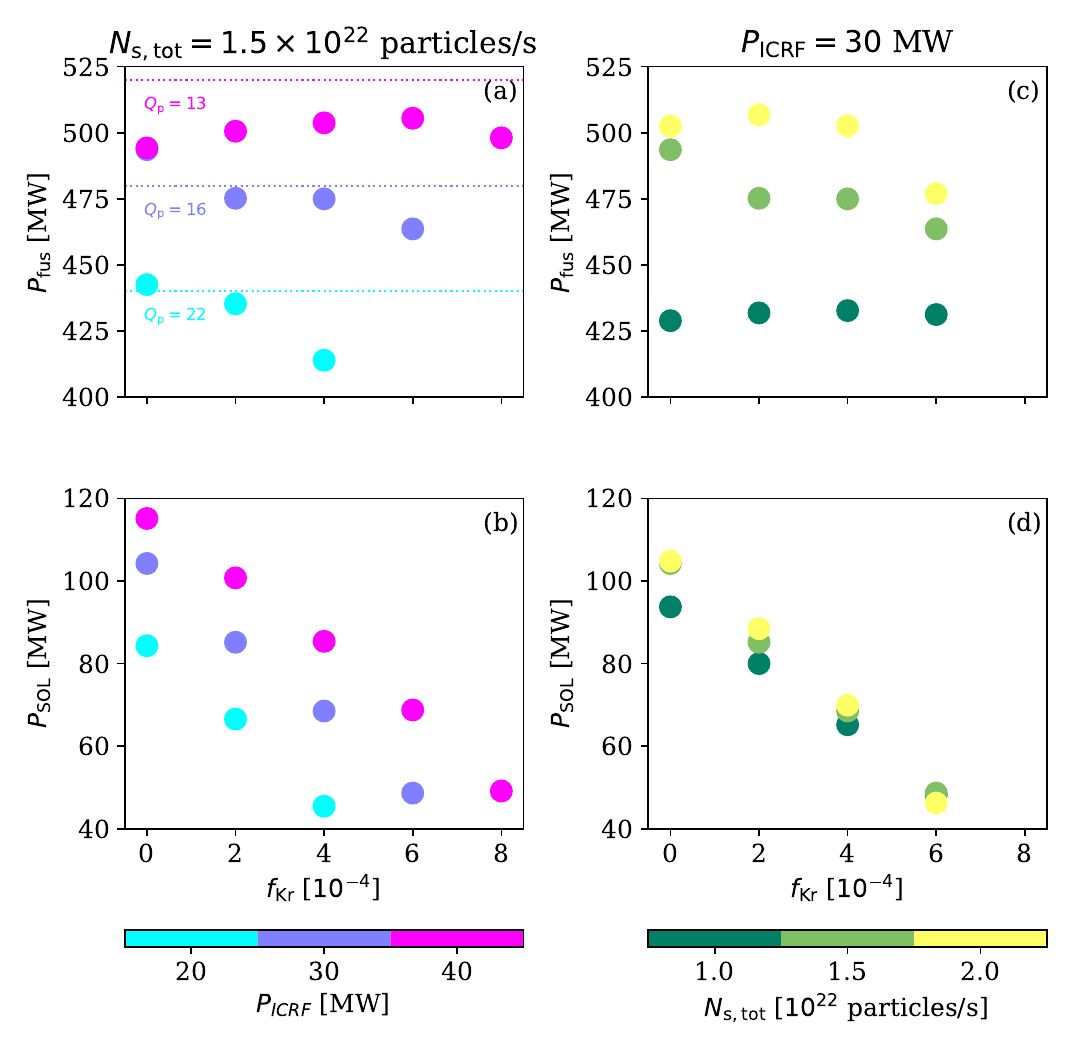}
    \caption{In plots (a) and (b), $\sub{P}{fus}$ and $\sub{P}{SOL}$, respectively, are plotted against $\sub{f}{Kr}$ at a constant $\sub{N}{s,tot} = 1.5\times 10^{22}$ particles per second. The color bar below these two plots gives $\sub{P}{ICRF}$, which was scanned at each $\sub{f}{Kr}$. Colored dotted lines at select $\sub{P}{fus}$ are labeled with the plasma gain of the $\sub{P}{ICRF}$ value corresponding to the color. In plots (c) and (d), $\sub{P}{fus}$ and $\sub{P}{SOL}$, respectively, are plotted against $\sub{f}{Kr}$ at a constant $\sub{P}{ICRF} = 30$ MW. The color bar below these two plots gives $\sub{N}{s,tot}$, which was scanned at each $\sub{f}{Kr}$.}
    \label{fig:pauxscan}
\end{figure*}

The gray 'Angioni' region in plot \ref{fig:peaking-c} is the range of predicted peaking values from equation \ref{eq:Angioni} for the simulations shown (including uncertainty from the scaling). The parameter $\sub{\Gamma}{NBI}^*$ is set equal to 0 because there is no NBI fueling. There is particle flux from pellet fueling, and reference \cite{angioniScalingDensityPeaking2007} mentions there is a similar relation between $\sub{pk}{Angioni}$ and fueling source peaking. However, because $\sub{\Gamma}{NBI}^*$ is strictly positive, including a non-zero value would only increase the predicted density peaking, making the agreement with the simulations here even worse. This implies the relation to fueling source peaking needs to be re-evaluated for sources that are further toward the edge, as they are depositing particles in regions of lower temperature and thus higher collisionality than on-axis sources. In general, the peaking matches prediction better as $\sub{\rho}{s}$ decreases and $\sub{N}{s,tot}$ increases. It is of note that the lowest density peaking shown here is 1.16, and all linear or logarithmic regressions that include collisionality in reference \cite{angioniScalingDensityPeaking2007} predicted the peaking of the ITER standard scenario \cite{mukhovatovComparisonITERPerformance2003} to be above 1.35. However, it was found in medium-fidelity transport studies of both ARC \cite{howardPerformanceTransportARC2026} and EU-DEMO \cite{fableRoleSourceCollisionality2019} that density peaking calculated by TGLF for these two FPP-class scenarios was below that predicted by empirical scalings. In the case of ARC, high-fidelity gyrokinetic modeling calculated an even lower density peaking than TGLF. However, neither of these cases included a fueling source, and it is expected that pellet fueling will only increase density peaking \cite{fableRoleSourceCollisionality2019} and therefore fusion performance.

We note that the source at $\rho = 0.5$ of width $\Delta_\mathrm{s} = 0.2\rho$ and $\sub{N}{s,tot} = 2.0 \times 10^{22}$ particles per second results in the highest $\sub{P}{fus}$ (605 MW). The $\sub{P}{fus}$ correlates almost directly to $pk_{\sub{n}{e}}$, with the lowest $\sub{P}{fus}$ (316 MW) at the lowest $pk_{\sub{n}{e}}$. This means $\sub{P}{fus}$ is similarly dependent on $\sub{\rho}{s}$ and $\sub{N}{s,tot}$. The total integrated particles of a source are dependent on the size of the pellet injected, which is thus a tunable experimental parameter.

For the remainder of this paper, we will use sources at $\rho = 0.6$, which is shown to be accessible in figure \ref{fig:pellet_deposition} and matches that found in pellet ablation models for CFETR \cite{mcclenaghanSelfconsistentInvestigationDensity2023,zhangSimulationStudyFueling2025}. The fuel source $\sub{N}{s,tot}$ will be selected based on what leads to convergence of the simulation since it is dependent on the pellet size, and the width of the Gaussian will be kept at $0.1\rho$ unless otherwise noted since it has a smaller effect on the output density profile.

\section{Potential for fusion gain improvement}
\label{sec:pauxscan}

In reference 
\cite{wilsonCharacterizingNegativeTriangularity2024}, scans of ICRF input power show a potential for increased fusion gain in a MANTA-like device by decreasing injected power.  However, this study was done by scaling a TORIC-calculated power deposition profile by a constant factor instead of self-consistently calculating power deposition profiles in TORIC for each injected power value. This is a good first approximation as the ICRF resonant location should not change significantly between simulations as it is dependent on magnetic field. However, the heat deposition profile is expected to change based on the temperature and density kinetic profiles, so self-consistent evolution with TORIC increases fidelity. Additionally, electron density was not evolved self-consistently and instead the scan utilized a scheme that maintained the density peaking always at the Angioni scaling prediction \cite{angioniScalingDensityPeaking2007}. 

In figure \ref{fig:pauxscan}, the result of self-consistently evolving auxiliary ICRH power with electron density, electron temperature, and ion temperature evolution in IPS-FASTRAN is shown. Each point on the plots represents a converged simulation. Scans were performed over the input ICRF power $\sub{P}{ICRF}$ from 20 MW to 40 MW, over $\sub{f}{Kr}$ from 0 to 0.0008, and over $\sub{N}{s,tot}$ from $1.0\times 10^{22}$ to $2.0 \times 10^{22}$ total integrated particles. The fueling source is a Gaussian centered at $\rho = 0.6$ with a width of $0.1\rho$. In plots \ref{fig:pauxscan-a} and \ref{fig:pauxscan-b}, $\sub{N}{s,tot}$ is at a constant $1.5\times 10^{22}$ particles per second and $\sub{P}{ICRF}$ is given by the colorbar below these two plots. In plots \ref{fig:pauxscan-c} and \ref{fig:pauxscan-d}, $\sub{P}{ICRF}$ is at a constant 30 MW, and $\sub{N}{s,tot}$ is given by the colorbar below these two plots. In plot \ref{fig:pauxscan-a}, colored dotted lines at select $\sub{P}{fus}$ are labeled with the value of $\sub{Q}{plasma}$ for the corresponding $\sub{P}{ICRF}$. This is to show the same trend here as was seen in reference \cite{wilsonCharacterizingNegativeTriangularity2024}, that plasma gain $\sub{Q}{plasma}$ can be increased by going to lower input power. The increase in gain at lower input powers is likely due to transport stiffness, that is, the MANTA operating point is above a critical gradient such that when heat flux is increased, the level of transport increases, so increasing input power has a lesser effect on the temperature profile \cite{dimitsComparisonsPhysicsBasis2000,ryterExperimentalEvidenceGradient2001,hillesheimExperimentalCharacterizationMultiscale2013, sauterNonstiffnessEdgeTransport2014}. The same transport stiffness trend was observed in high-fidelity gyrokinetic modeling of a high-field L-mode tokamak scenario in reference \cite{rodriguez-fernandezCorePerformancePredictions2024}. However, the total fusion power and auxiliary power are what determine the net electric power $\sub{P}{e,net}$. According to the super-critical Rankine cycle model used in the original MANTA work \cite{collaborationMANTANegativetriangularityNASEMcompliant2024}, $\sub{P}{fus}$ must reach more than $300$ MW to generate at least $\sub{P}{e,net} = 50$ MWe with $\sub{P}{aux} = 40$ MW. If $\sub{P}{aux}$ decreases, $\sub{P}{e,net}$ only increases, so every simulation shown in figure \ref{fig:pauxscan} meets $\sub{P}{e,net} > 50$ MWe.

\begin{figure*} 
    \centering
    \includegraphics[width=\linewidth]{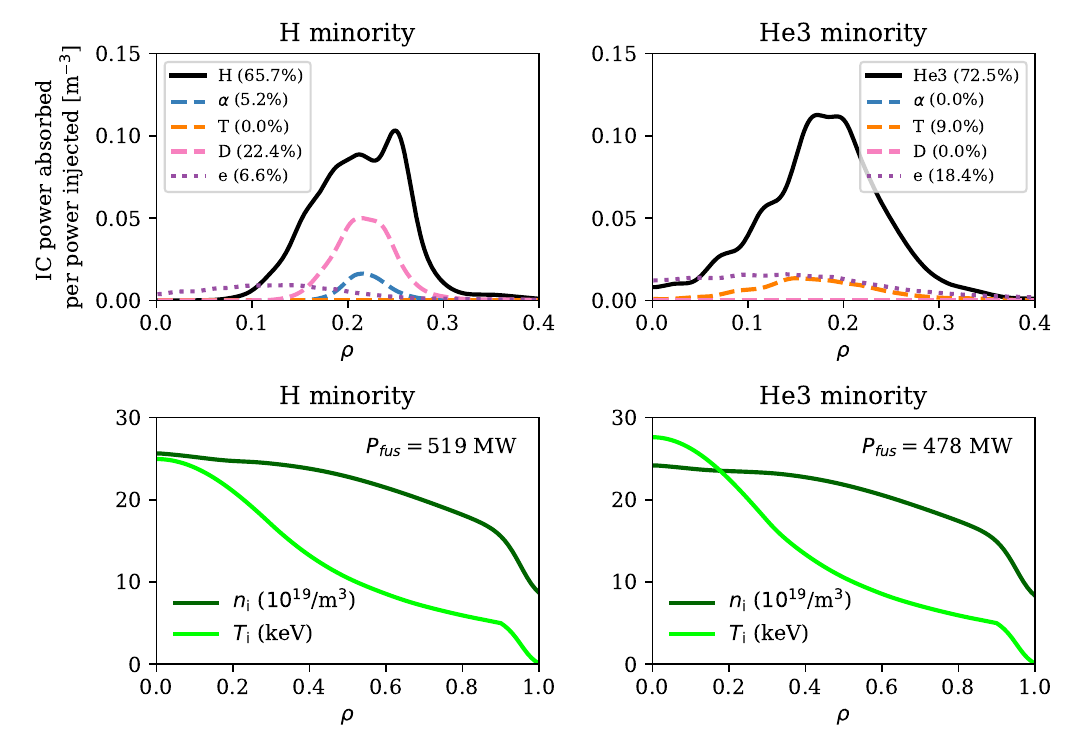}
    \caption{Select profiles comparing the use of hydrogen and helium-3 as minority ions, with a launched wave frequency of 170 MHz and 110 MHz, respectively, to obtain matching resonance layer locations. In the top plots, the IC power absorption per megawatt power injected of the minority ion (black), alphas (red), tritium (blue), and deuterium (purple) is shown for hydrogen as a minority ion on the left and helium-3 as a minority ion on the right. Solid lines represent absorption at the fundamental resonance of the labeled species, while dashed lines are at the second harmonic resonance. The electron absorption is shown by the dotted line. The percentage of power absorbed by each species is indicated in the legend. Only the species and resonances with the largest power absorbed are shown. In the bottom plots, ion temperature and ion density are plotted versus normalized radius in light green and dark green, respectively.}
    \label{fig:He3versusH}
\end{figure*}

It is of note that in the $\sub{f}{Kr} = 0.0006$ case, the simulations with 20 MW of input power did not converge. This is likely because the power from fusion and ICRF heating is not enough to compensate for the power loss from line radiation. In general, reducing the krypton fraction results in higher $\sub{Q}{plasma}$ but also higher $\sub{P}{SOL}$. Increasing fueling magnitude increases $\sub{P}{fus}$ while minimally increasing $\sub{P}{SOL}$, implying a larger source is desirable, which would correspond to a larger fuel pellet in practice. We note that difficulties in fabricating a larger fuel pellet and the monetary cost of a larger fuel source would need to be taken into consideration as well if this path were pursued. 

The results shown in figure \ref{fig:pauxscan} are for our base case with current, temperature, and density evolution. The only change to the heating scheme from the MANTA-1 case is the magnitude of injected ICRF power, with heating evolved in TORIC. We will now discuss potential optimization of the H\&CD scheme that could further improve the performance of the base case besides going to lower input power.

\section{IC heating optimization}
\label{sec:ic}

Ion cyclotron (IC) resonant heating is currently the best candidate for on-axis heating in a high-field tokamak \cite{linPhysicsBasisICRF2020}. Electron cyclotron (EC) resonant heating on-axis in a device with an on-axis magnetic field greater than $\sim$8T would require launched wave frequencies above those which modern gyrotrons can produce \cite{shinyaDemonstrationHighpowerGyrotron2026, linPhysicsBasisICRF2020}. Efficient on-axis IC heating requires including a low-Z minority ion in the plasma \cite{stixFASTWAVEHEATINGTWOCOMPONENT,frankSimulatingEnergeticIons2024}. Our primary analysis of IC resonant heating will be with helium-3 to match the original MANTA design, but we will also investigate using hydrogen-1 as it is more abundant and cost-effective. 

\begin{figure*} 
    \centering
    \includegraphics[width=0.49\linewidth]{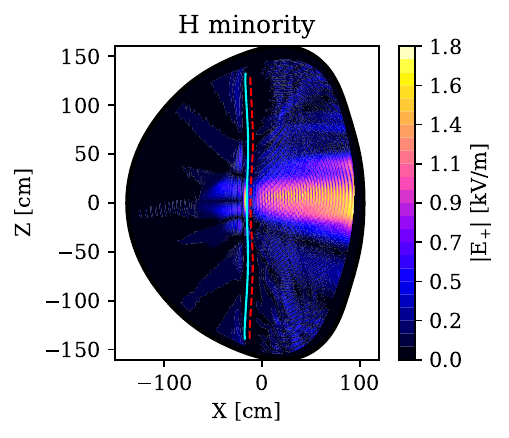}
    \includegraphics[width=0.49\linewidth]{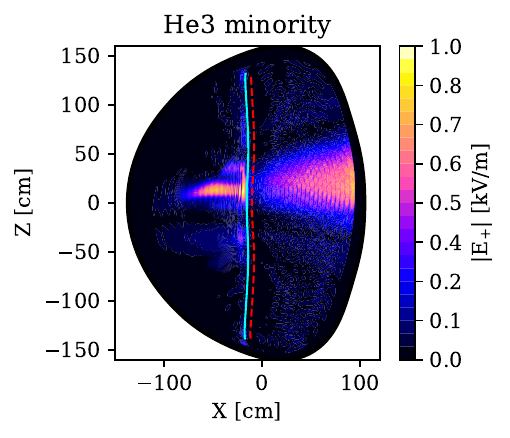}
    \caption{The normalized $\sub{E}{+}$ is plotted in arbitrary units in the base case simulation using hydrogen-1 (left) and helium-3 (right) as a minority ion calculated by TORIC. The resonance layer for each species is shown by the red dashed line. The ion-ion cutoff layer is shown by the solid blue line.}
    \label{fig:He3versusH2D}
\end{figure*}

\subsection{Minority ion options}
The location of the resonance layer on which an injected wave will damp is dependent on the ion mass $\sub{m}{i}$ and charge $Z$ based on the equation for ion cyclotron frequency $\sub{\omega}{c,i}=Z\sub{q}{e} B/\sub{m}{i}$. To solve for the magnetic field at which the injected wave will damp $\sub{B}{res}$, we rearrange the equation to be 
\begin{equation}
    \sub{B}{res} = \frac{\omega \sub{m}{i}}{Z\sub{q}{e}}
\end{equation}
where $\omega$ is the injected wave frequency. For a given injected wave frequency, we can then evaluate how $\sub{B}{res}$ is dependent on $Z$ and $\sub{m}{i}$ of a given ion. From this evaluation, hydrogen-1 and helium-3 seem to be on equal footing as an injected wave at a given frequency would damp on either of them before damping on alpha particles (helium-4). 

However, looking at the second harmonic resonance illuminates that the helium-4 $n=2$ resonant B-field is at the same value as the hydrogen-1 fundamental resonance. Therefore, targeting the hydrogen-1 fundamental resonance can result in inadvertent heating of alpha particles. An example of this is shown in figure \ref{fig:He3versusH}. To compare the two minority ion options, we ran a TORIC simulation with helium-3 and an injected wave at 110 MHz and a simulation with hydrogen-1 and an injected wave at 170 MHz, such that the resonance layer is at the same location in the plasma for both cases. Both minority ions are included at the same fraction ($\sub{f}{minority}  = 0.02$), and the electron density is held constant at the profile from the original MANTA operating point. In the top plots of figure \ref{fig:He3versusH}, we see that the hydrogen-1 minority plasma heats the second harmonic resonance of deuterium and alphas, while the helium-3 heats only the tritium second harmonic resonance significantly. For helium, the split of total absorbed power is 9\% to tritium, 73\% to helium-3, and 18\% to electrons. For hydrogen, the power split is 22\% to deuterium, 66\% to hydrogen-1, 5\% to alphas, and 7\% to electrons. Heating alphas is undesirable as it can lead to increased fast-ion losses, which result in higher wall heat loads and sputtering \cite{hillesheimOverviewPhysicsBasis2026}. The hydrogen-1 minority plasma results in a higher fusion power than the helium-3 minority plasma when electron density is held constant ($\sub{P}{fus} = 478$ MW for helium-3 versus $\sub{P}{fus} = 519$ MW for hydrogen-1, a 9\% increase). This is mostly due to the lower fuel dilution (hydrogen has lower Z), which is shown in the bottom plots of figure \ref{fig:He3versusH}, where $\sub{n}{i}$ is higher for the hydrogen-1 minority plasma, and even though $\sub{T}{i}$ is lower, this results in higher $\sub{P}{fus}$. This implies a similar increase in $\sub{P}{fus}$ could be reached by going to a lower helium-3 fraction, a possibility which will be discussed in subsection \ref{subsec:he3frac}.


In figure \ref{fig:He3versusH2D}, the left-hand circularly-polarized component of the normalized electric field $\sub{E}{+}$ is plotted in arbitrary units for helium-3 and hydrogen-1 as minority ions in the TORIC simulation. The wave is launched from the outboard midplane in both cases. In the helium-3 case, there is a significant amount of mode conversion seen at the ion-ion cutoff. In the hydrogen-1 case, there is very little mode conversion. The apparent increase in magnitude of the injected wave (indicated by the color bar in figure \ref{fig:He3versusH2D}) is likely due to weak coupling as the antenna loading is lower in the H minority case than in the He3 minority case. 

\subsection{High field side IC launch}
\begin{figure}
    \centering
    \includegraphics[width=\linewidth]{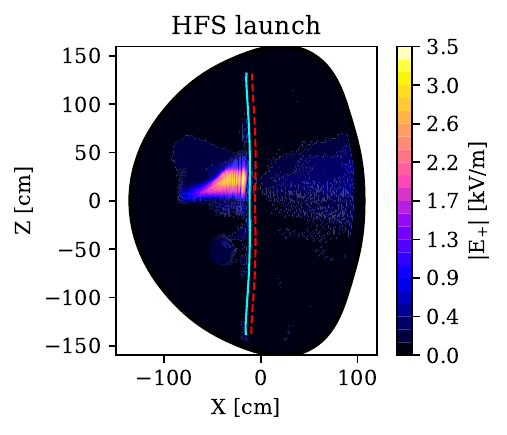}
    \caption{The normalized $\sub{E}{+}$ is plotted in arbitrary units for an ICRF wave launched $30^\circ$ above the inboard midplane calculated by TORIC. The ion-ion hybrid cutoff layer is indicated by the solid blue line, and the helium-3 resonance layer is indicated by the red dashed line.}
    \label{fig:HFS_IC}
\end{figure}

The location of the ion-ion hybrid cutoff on the inboard side of the resonance layer is an obstacle to efficient heating from high field side (HFS) launch of ICRF waves. An example of this is shown in figure \ref{fig:HFS_IC}. The inboard midplane is likely an inaccessible location for ICRF, as this is the area where the plasma is closest to the central solenoid. Thus, the wave is injected from $30^\circ$ above the inboard midplane. The injected wave reaches the ion-ion hybrid cutoff layer before the minority fundamental resonance layer when launched from the inboard side in this configuration. This results in a large amount of mode conversion, which in figure \ref{fig:HFS_IC} is made evident by the relative magnitude of $\sub{E}{+}$ of the reflected wave compared to the injected wave. The high mode conversion is also evident in the relative power deposition to ions and electrons, with high field side launch resulting in 22 MW of ion heating and 18 MW of electron heating. Because ion heating is preferential for increasing fusion power, inboard launch is not appealing for ICRF heating in this configuration. 

\begin{figure}
    \centering
    \labelphantom{fig:lownphi-a}
    \labelphantom{fig:lownphi-b}
    \labelphantom{fig:lownphi-c}
    \includegraphics[width=\linewidth]{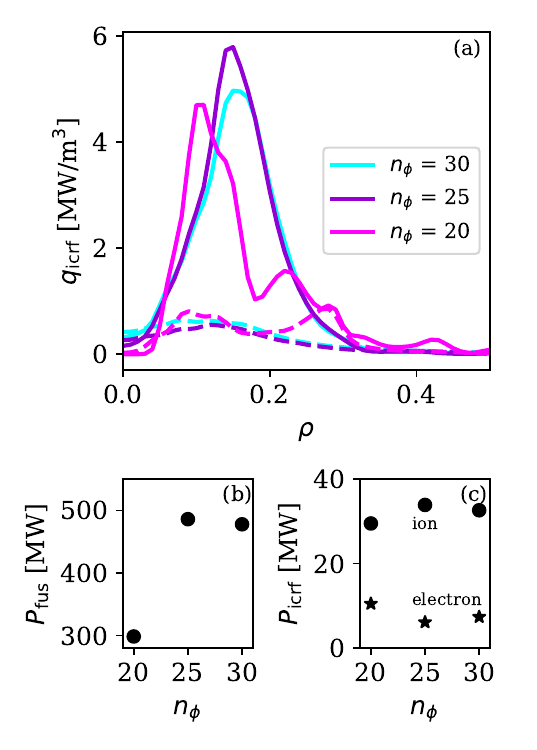}
    \caption{ICRF power density is given in (a), with pink, purple, and blue lines corresponding to $\sub{n}{\phi} =$ 20, 25, and 30, respectively. The solid and dashed lines indicate power to the ions and electrons, respectively. In (b) and (c), the fusion power and ICRF power are plotted versus $\sub{n}{\phi}$, respectively. In (c), dots indicate power to ions and stars indicate power to electrons.}
    \label{fig:lownphi}
\end{figure}

\subsection{Changing toroidal mode number}
The level of mode conversion in the helium-3 case can be reduced by slightly lowering the toroidal mode number $\sub{n}{\phi}$ of the injected wave. Lowering $\sub{n}{\phi}$ from 30 to 20 outputs the results shown in figure \ref{fig:lownphi} for the helium-3 case. The lower ICRF power to electrons $\sub{q}{e,icrf}$ is due to lower mode conversion between the $\sub{n}{\phi} = 30$ and $\sub{n}{\phi} = 25$ cases. Lower mode conversion from this change leads to an increase in fusion power of only about 2\% due to the increased ion heating. However, further lowering $\sub{n}{\phi}$ to 20 results in a drop in $\sub{P}{fus}$ of almost 40\% from the $\sub{n}{\phi} = 30$ case. This is due to increased mode conversion, as evidenced by the increase in ICRF power to electrons. The change to increased mode conversion is likely due to a transition from the "minority heating regime" to the "mode conversion regime." This transition is discussed in reference \cite{kazakovEffectImpuritiesTransition2013}, where it is shown that lowering $\sub{n}{\phi}$ decreases the Doppler width of the minority IC resonance and thus decreases the helium-3 impurity fraction at which the transition occurs. In the mode conversion regime, the injected fast wave is partially converted to the ion Bernstein wave and ion cyclotron wave at the mode conversion layer, and these short wavelength modes are strongly absorbed by electrons \cite{kazakovEffectImpuritiesTransition2013}. Thus, while $\sub{n}{\phi} = 25$ increases $\sub{P}{fus}$ slightly, it is not significant enough to prioritize changing the original MANTA antenna design with the same minority fraction. However, reducing the minority fraction could increase $\sub{P}{fus}$ from lower dilution as well as avoid a transition to the mode conversion regime, an option we will discuss in the next subsection. Generally, a lower $\sub{k}{\parallel}$ requires a larger inter-strap distance of the antenna \cite{usoltcevaDEMOIonCyclotron2021}. We note that the EU-DEMO antenna design has targeted a parallel wavenumber $\sub{k}{\parallel}$ of 4-5 m$^{-1}$ \cite{usoltcevaDEMOIonCyclotron2021} and that a $\sub{k}{\parallel} = 6.4$ m$^{-1}$ is considered feasible by ARC \cite{hillesheimOverviewPhysicsBasis2026}. By the approximate relation at the outboard antenna $\sub{k}{\parallel} \approx \sub{n}{\phi}/R$, $\sub{n}{\phi} = 25$ results in a $\sub{k}{\parallel}$ of about 5.5 m$^{-1}$.



\subsection{Fraction of minority ion}
\label{subsec:he3frac}
Another factor influencing the efficiency of ICRF is the fraction of the minority ion included. The fraction must be optimized to maximize antenna loading and ensure single-pass damping, but minimize dilution, which negatively impacts performance. Higher minority ion fractions result in both higher fuel dilution and dramatically increased mode conversion past a critical value (a transition from the minority heating regime to the mode conversion regime), which results in more electron heating and a reduction of single-pass absorption \cite{kazakovEffectImpuritiesTransition2013}. However, if the minority fraction is not high enough, then it can lead to poor antenna coupling, that is, less power absorbed by the plasma for the same amount of current in the antenna straps. 

\begin{figure}
    \centering
    \labelphantom{fig:he3fracscan-a}
    \labelphantom{fig:he3fracscan-b}
    \labelphantom{fig:he3fracscan-c}
    \labelphantom{fig:he3fracscan-d}
    \includegraphics[width=\linewidth]{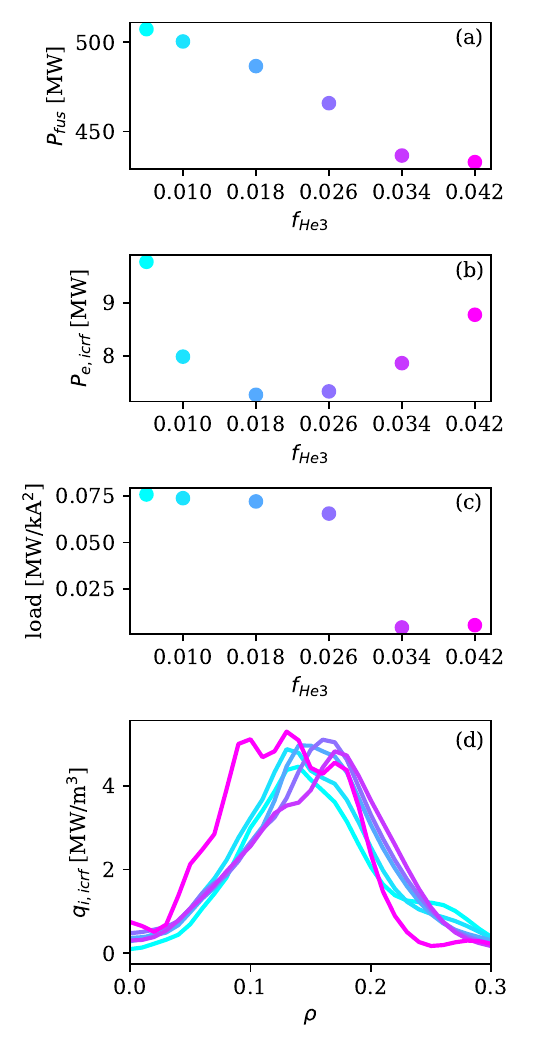}
    \caption{In this figure, the points are colored for each helium-3 fraction to correspond to the colors of profiles in plot (d). In plots (a), (b), and (c), the fusion power, ICRF power to electrons, and antenna load, respectively, are plotted against helium-3 fraction. In (d), the power density from ICRF to the ions is plotted against normalized minor radius from $\rho = 0$ to $\rho = 0.3$.}
    \label{fig:he3fracscan}
\end{figure}

In figure \ref{fig:he3fracscan}, a scan of the helium-3 fraction $\sub{f}{He3}$ with a constant electron density profile is shown. Ion density is calculated by assuming quasineutrality and changes only slightly from varying $\sub{f}{He3}$ from 0.006 to 0.042. The antenna loading is almost constant for lower $\sub{f}{He3}$ values scanned, under $\sub{f}{He3} = 0.026$. At $\sub{f}{He3} = 0.034$, it is likely that a transition to the mode conversion regime occurred, as there is a jump in electron heating (seen in figure \ref{fig:he3fracscan-b}) and $|\sub{E}{+}|$ is localized at the ion-ion hybrid cutoff. This likely contributes to the reduced $\sub{P}{fus}$ as well. The decrease in $\sub{P}{fus}$ going to higher $\sub{f}{He3}$ is mainly attributable to the increased dilution. While a lower $\sub{f}{He3}$ is desirable for increased fusion power, too low a minority fraction reduces single-pass absorption \cite{kazakovEffectImpuritiesTransition2013}. We note that a high level of impurity fraction control in a real experiment would be difficult, so maintaining the minority fraction between 0.015 and 0.03 would be optimal for avoiding the mode conversion regime while maintaining good absorption.

\begin{figure*}
    \centering
    \includegraphics[width=\linewidth]{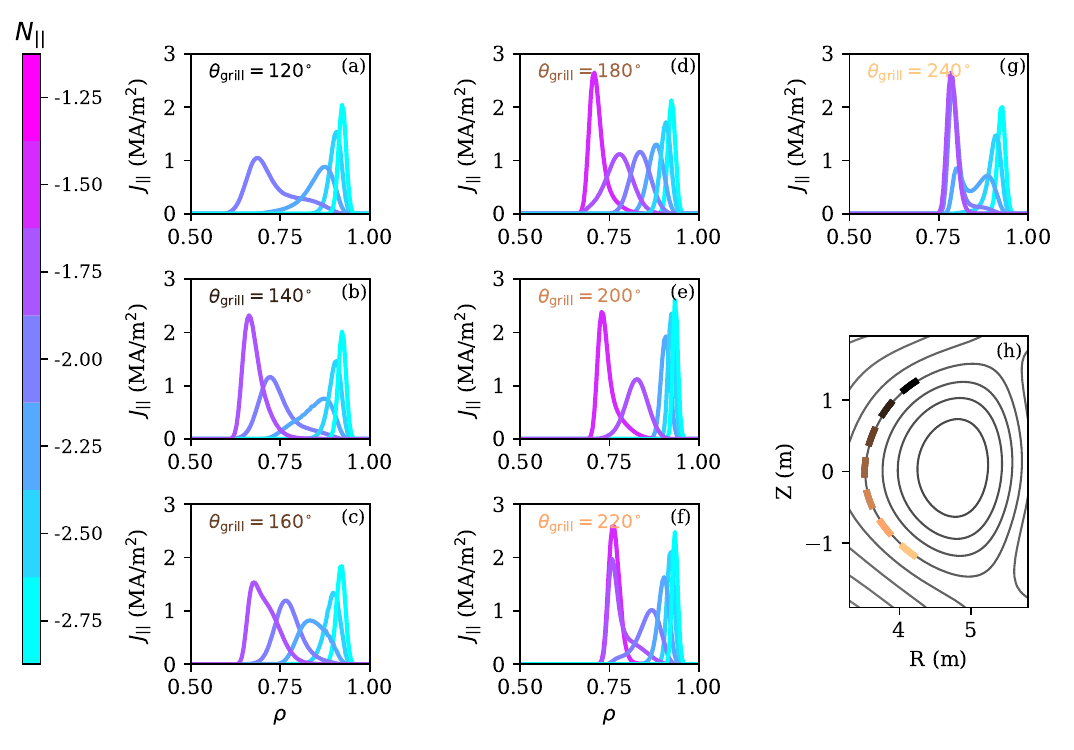}
    \caption{Current profiles from LH waves launched at seven distinct poloidal locations ($\sub{\theta}{grill}$) and seven distinct parallel refractive indices ($\sub{N}{\parallel}$) are shown, with each plot corresponding to a different $\sub{\theta}{grill}$ and colorbar indicating $\sub{N}{\parallel}$. The 2D flux surface plot (h) shows all the launch locations on the equilibrium.}
    \label{fig:LHcurrentprofs}
\end{figure*}

\begin{figure*}
    \centering
    \includegraphics[width=\linewidth]{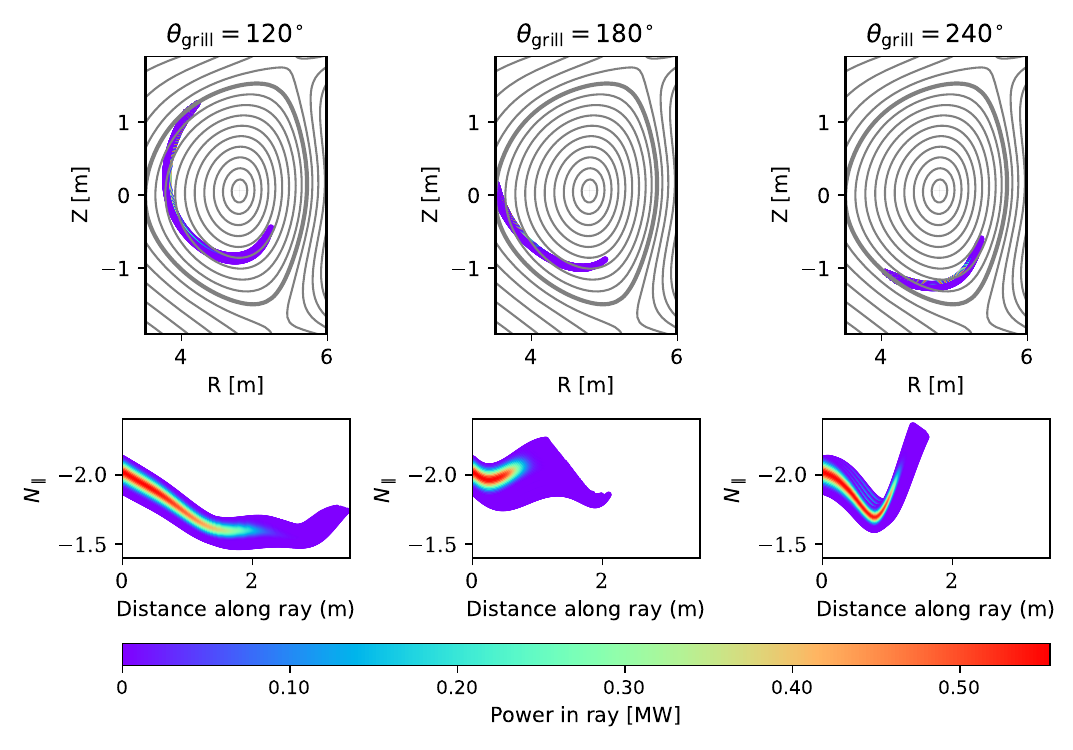}
    \caption{Waveforms for LH launches at three distinct poloidal locations with $\sub{N}{\parallel} = -2.0$ are shown. The bottom plots show the change in $\sub{N}{\parallel}$ against the distance along the ray. The colorbar indicates the power in the ray.}
    \label{fig:LHwaveforms}
\end{figure*}

\section{LH current drive accessibility}
\label{sec:lh}

Lower hybrid current drive (LHCD) is appealing for use in FPP-class devices as it is one of the most efficient current drive methods and works in high-field applications. However, it is more difficult to predict where the wave will damp as the lower hybrid frequency is dependent on both density and magnetic field, so the location of wave launch is more consequential on the current profile than for cyclotron resonance H\&CD methods. 

\subsection{The benefits of inboard launch}

In an NT device, there is more inboard space available than in a PT device, where the majority of the plasma is on the high-field side. This allows greater opportunity for inboard side launch. Inboard side launch is appealing for all wave launchers from an engineering perspective as it can increase the lifetime of the launcher because of decreased heat flux on the inboard side \cite{bonoliHighFieldSide2018,rutherfordEvaluationDIIIDPlasmas2026,rutherfordPredictionsHighField2026}. From a physics perspective, it is appealing for LHCD because it provides a larger window of propagating slow waves and access to a double-mode conversion upshift which can increase the current drive efficiency \cite{bonoliHighFieldSide2018,rutherfordOptimizationN_parallelUpshift2024}. Also, the increased SOL quiescence could mean fewer losses from parametric decay instabilities and scattering of LH waves from density fluctuations \cite{bonoliHighFieldSide2018}. Previous work has also shown that in a high-field NT FPP-class configuration, NT plasmas allow for a broader range of launcher locations for single-pass absorption \cite{rutherfordPredictionsHighField2026}. 

\subsection{Accessible LH current profiles and stabilization potential}

The wave damping location in LHCD is dependent on density as well as magnetic field. Due to the high density of FPP-class plasmas, penetration into the core is not possible. Changing the launch angle $\sub{\theta}{grill}$ (measured from the outboard midplane and conformal to the SOL) and parallel refractive index $\sub{N}{\parallel}$ changes the current deposition location \cite{hassanCorePedestalPlasmaConfigurations2023,rutherfordPredictionsHighField2026}. A scan of these two variables is shown in figure \ref{fig:LHcurrentprofs}. Only profiles resulting from simulations with single-pass absorption are shown. The launched power for all profiles is 25 MW. The $\sub{\theta}{grill}$ value is written at the top of each current profile plot, and the color of the text corresponds to the launcher locations shown on the equilibrium in the bottom right plot. The color bar gives $\sub{N}{\parallel}$. In general, the current profile peak moves toward larger $\rho$ as $\sub{N}{\parallel}$ becomes more negative. The upper launch locations are able to drive current at smaller $\rho$. The farthest inward current profile driven is in the case with $\sub{\theta}{grill} = 140^\circ$ and $\sub{N}{\parallel} = -1.75$, with a peak around $\rho = 0.65$. The largest total driven current of the profiles shown in figure \ref{fig:LHcurrentprofs} is 1.2 MA (for the $\sub{N}{\parallel} = -1.75$ and $\sub{\theta}{grill} = 140^\circ$ case), which would constitute about 12\% of the total current in the MANTA configuration. 

In figure \ref{fig:LHwaveforms}, the rays for three select cases of LH launch are shown along with their launched $\sub{N}{\parallel}$ values and how they change along the ray. All rays shown are launched with $\sub{N}{\parallel} = -2.0$. The color bar gives the power in the ray, reaching zero as the power is fully deposited. The left, middle, and right plots are from cases launched at $\sub{\theta}{grill} = 120^\circ$, $180^\circ$, and $240^\circ$, respectively. This figure once again demonstrates that waves launched from the upper half of the plasma result in current driven further inward, as the wave reaches further into the plasma before completely damping. We also observe that the wave is able to propagate farther at less negative $\sub{N}{\parallel}$, which matches the trend seen in figure \ref{fig:LHcurrentprofs} where less negative $\sub{N}{\parallel}$ resulted in current deposition further inward.

On-axis current drive is desirable for maintaining confinement \cite{KnovelMagnetohydrodynamics} and controlling the central safety factor value \cite{driveChapter6Plasma1999}. At the high density and magnetic field of an FPP-class device, LHCD cannot drive current on-axis. However, off-axis current drive allows current profile tailoring and is useful for certain stabilization applications. For example, additional current drive at the $q=2$ surface can stabilize neoclassical tearing modes \cite{woodbyModelCurrentDrive2008} and current driven in the edge can counteract vertical instability driven by a drop in edge current density \cite{tobinRealtimeObservationToroidal2026a}. In the base case equilibrium, the $q=2$ surface is located far into the edge (at about $\rho = 0.95$), which, given the spread of the current deposition profiles, would result in scrape-off layer interactions if the LH current were driven at this location. Therefore, we ran the same LH configuration on an altered equilibrium with a higher $q$ to test access of the LH driven current to the $q=2$ surface. 

\begin{figure}
    \centering
    \includegraphics[width=\linewidth]{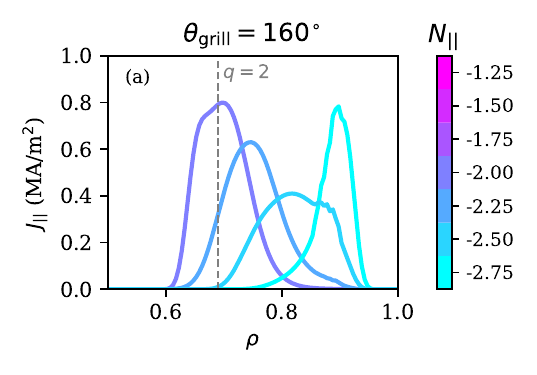}
    \includegraphics[width=\linewidth]{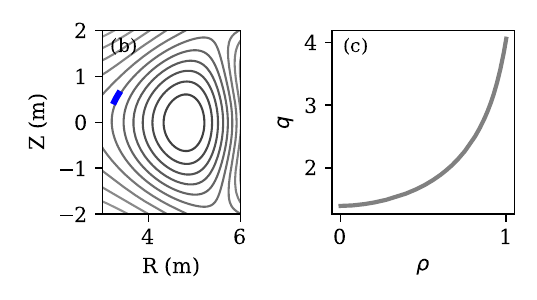}
    \caption{In (a), the current density profiles for LH launch on a higher $q$ equilibrium are shown, with the colorbar indicating $\sub{N}{\parallel}$ and the $q=2$ surface marked by the dashed gray line. In (b), the modified equilibrium is shown with the blue rectangle indicating the launcher location chosen for the current density localized at the $q=2$ surface. The $q$ profile for this equilibrium is shown in (c).}
    \label{fig:LH_newq}
\end{figure}

The results of LH launch on the higher $q$ equilibrium are shown in figure \ref{fig:LH_newq} for $\sub{\theta}{grill} = 160^\circ$. The $q=2$ surface is located at about $\rho = 0.7$, which is only accessible by waves launched above the inboard midplane in this configuration, as seen in figure \ref{fig:LHcurrentprofs}. A narrow current density profile is desirable for stabilization of neoclassical tearing modes \cite{rutherfordPredictionsHighField2026,hayeCrossMachineBenchmarking2006,reimanSuppressionTearingModes2018}, and the launch location of $\sub{\theta}{grill} = 160^\circ$ resulted in the narrowest current density peak at the $q=2$ surface. Waves launched with $|\sub{N}{\parallel}|>2$ were not single-pass and thus are not shown. Comparing the current profiles in figure \ref{fig:LH_newq} with those at the same $\sub{\theta}{grill}$ value in figure \ref{fig:LHcurrentprofs}, it is clear that the change in equilibrium significantly affects the current deposition, which highlights the importance of integrated modeling for LHCD analysis as the LH configuration must be optimized in conjunction with the equilibrium and pressure profiles.

\section{Conclusion}
\label{sec:conclusion}
In this work, MANTA \cite{collaborationMANTANegativetriangularityNASEMcompliant2024,millerPowerHandlingHighlyradiative2024,wilsonCharacterizingNegativeTriangularity2024} is implemented in the IPS-FASTRAN framework \cite{parkIntegratedModelingHigh2018} for fueling, heating, and current drive studies. We scanned a variety of Gaussian pellet-like fueling sources to be used in simulations with density evolution for more accurate performance prediction. In these scans, we found that the radial deposition location of the fuel source influences performance more than the deposition width of the source. Additionally, the empirical scaling in \cite{angioniScalingDensityPeaking2007} was found to over-predict density peaking, matching the conclusion drawn from recent results from the high-field tokamak fusion pilot plant design ARC \cite{howardPerformanceTransportARC2026}. A simple pellet ablation model \cite{mcclenaghanSelfconsistentInvestigationDensity2023} was used to verify the feasibility of the fuel sources used for the remainder of the paper. The inboard space that NT affords can be used for high-speed high field side launch of pellets, taking advantage of the beneficial drift direction on the high-field side. 

Ion cyclotron resonant frequency (ICRF) heating is the best candidate for on-axis heating in a high-field device \cite{linPhysicsBasisICRF2020}. Choosing hydrogen-1 as the minority ion yielded higher fusion power from reduced fuel dilution and reduced mode conversion, but has the drawback of heating alphas as well, which could lead to increased fast ion losses. Dedicated studies on whether the increased fusion power and affordability of hydrogen-1 outweigh these losses are encouraged. 

Lowering the toroidal mode number of the launched wave for helium-3 minority ICRF reduced mode conversion at first (matching results found in \cite{frankSimulatingEnergeticIons2024}) and resulted in an increase in fusion power of only about 2\%. However, further reduction of the toroidal mode number led to a transition to the mode conversion regime and a significant loss of fusion power from decreased ion heating. 

Varying the helium-3 fraction results in a slightly larger change in fusion power, increasing fusion power by about $\sim 8$\% with a decrease in helium-3 fraction of about $\sim 60$\%. Increasing helium-3 fraction past a critical point results in a transition from the resonant heating regime to the mode conversion regime. Thus, operating at lower helium-3 fractions is desirable as long as good single-pass absorption and antenna coupling are maintained. Lowering the input power from ICRF from 40 MW to 20 MW can increase fusion gain and lower scrape-off layer power. Seeded krypton is used to lower scrape-off layer power as well, but lowering the input power while increasing krypton fraction can lead to radiative collapse due to the increased radiative power. 

The inboard space afforded by NT geometry grants the opportunity for inboard launch of lower hybrid (LH) waves. In the MANTA configuration, varying the parallel refractive index of the launched wave from -2.75 to -1.25 and the poloidal location (measured from the outboard midplane) from 120$^\circ$ to 240$^\circ$ results in current profiles from single-pass absorption that peak in the edge region between $\rho = 0.6$ and $0.9$, generating up to 1.2 MA of current. It is important to evaluate the effects of LH current drive (LHCD) with integrated modeling as the resulting current density is greatly affected by the equilibrium and pressure profile as well as the choice of launcher location and parallel refractive index. Future work integrating the interplay of LHCD and ICRF heating with a higher fidelity model for LHCD is essential for continued study of the effectiveness of LHCD in a high-field fusion pilot plant.
    
\section*{Acknowledgments}
 This work was supported by the U.S. Department of Energy, Office of Science, Office of Fusion Energy Sciences under Award DE-SC0022272, and based upon work supported by the U.S. Department of Energy, Office of Science, Office of Workforce Development for Teachers and Scientists, Office of Science Graduate Student Research (SCGSR) program. The SCGSR program is administered by the Oak Ridge Institute for Science and Education (ORISE) for the DOE. ORISE is managed by ORAU under contract number DE-SC0014664. All opinions expressed in this paper are the authors' and do not necessarily reflect the policies and views of DOE, ORAU, or ORISE.

\section*{References}

\bibliographystyle{iopart-num}
\bibliography{references.bib}


\end{document}